\documentclass[
 twocolumn,
 groupedaddress,
 showpacs,
 amsmath,
 amssymb,
 aps,
 pre
]{revtex4-1}

\usepackage[dvipdfmx]{graphicx}
\usepackage{dcolumn}
\usepackage{bm}

\usepackage{algcompatible}
\usepackage[size=small]{caption}
\DeclareCaptionFormat{algorithm}{\hrulefill\vskip-8pt\hrulefill\par#1#2#3\vskip-4pt\hrulefill}
\usepackage{etoolbox}
\AtEndEnvironment{algorithm}{\noindent\hrulefill\par\nobreak\vskip-8pt\hrulefill}
\usepackage{newfloat}
\DeclareFloatingEnvironment[
  fileext=loa,
  listname=List of Algorithms,
  name=ALGORITHM,
  placement=tbhp,
]{algorithm}

\usepackage{mathtools}
\usepackage{amsthm}

\usepackage{subcaption}
\usepackage{xcolor}
\usepackage{diagbox}
\usepackage{multirow}
\makeatletter
\newcommand*\bigcdot{\mathpalette\bigcdot@{.5}}
\newcommand*\bigcdot@[2]{\mathbin{\vcenter{\hbox{\scalebox{#2}{$\m@th#1\bullet$}}}}}
\makeatother
\usepackage{xr}
\usepackage{braket}
\usepackage{amsmath}

\makeatletter
\newcommand{\vast}{\bBigg@{3.0}}
\newcommand{\Vast}{\bBigg@{3.5}}

\makeatother

\begin{document}

\preprint{APS/123-QED}

\title{Work relations with measurement and feedback control on nonuniform temperature systems}

\author{Hideyuki Miyahara}
\email{hideyuki\_miyahara@mist.i.u-tokyo.ac.jp}

\affiliation{%
Department of Mathematical Informatics,
Graduate School of Information Science and Technology,
The University of Tokyo,
7-3-1 Hongo, Bunkyo-ku, Tokyo 113-8656, Japan
}%

\author{Kazuyuki Aihara}
\affiliation{%
Institute of Industrial Science, The University of Tokyo,
4-6-1 Komaba, Meguro-ku, Tokyo 153-8505, Japan
}%
\affiliation{International Research Center for Neurointelligence (WPI-IRCN), UTIAS,
The University of Tokyo,
7-3-1 Hongo, Bunkyo-ku, Tokyo 113-0033, Japan}

\date{\today}


\begin{abstract}
The relation between the work performed to a system and the change of its free energy during a certain process is important in nonequilibrium statistical mechanics.
In particular, the work relation with measurement and feedback control has attracted much attention, because it resolved the paradox concerning Maxwell's demon.
Most studies, however, assume that their target systems are isolated or isothermal.
In this paper, by considering a nonisothermal system, we generalize the Sagawa-Ueda-Jarzynski relation, which involves measurement and feedback control, and apply it to a realistic model.
Furthermore, when the temperature profile is quadratic, we see that the system is governed by Tsallis statistical mechanics.
In addition, we show that our formulation provides the generalized version of the second law of information thermodynamics and a set of new work relations for isothermal systems.
\end{abstract}


\maketitle


\section{Introduction}

Stochastic thermodynamics has been intensively studied in last two decades~\cite{Jarzynski02, Seifert01, Seifert02, Seifert03, Sekimoto01, Seifert04, Kurchan01}, and some notable relations between the work performed to a system and its free energy change during a nonequilibrium process, such as the Jarzynski equality~\cite{Jarzynski01, Jarzynski03} and the Crooks relation~\cite{Crooks01}, were found.
Subsequently, statistical mechanics and information were combined in Ref.~\cite{Sagawa01, Sagawa02, Sagawa03, Sagawa04, Sagawa05, Parrondo01}, and the Jarzynski equality was generalized for a system that involves measurement and feedback control.
These works are basically based on thermodynamics and Boltzmann statistical mechanics.

On the other hand, multiplicative noise also ubiquitously appears in physics and other fields of science, and then its thermodynamic properties are of considerable interest~\cite{Matsuo01}.
One of the most interesting characteristics of stochastic dynamics with multiplicative noise is that it leads to an anomalous diffusion process, which has a close relation with Tsallis statistical mechanics~\cite{Tsallis01, Ford01}.
From the viewpoint of stochastic thermodynamics, the Jarzynski equality was extended on the basis of a nonuniform temperature system and Tsallis statistical mechanics~\cite{Ford01, Ponmurugan01}.

In this paper, we provide the work relations under measurement and feedback control for a nonisothermal system by generalizing the relation shown in Ref.~\cite{Sagawa03}.
This problem setting appeared in Ref.~\cite{Ford01}, but, in this paper, we generalize it to the case where the confining potential includes measurement and feedback control.
Then, we discuss the generalized second law of information thermodynamics of our framework.
We see that the generalized second law of information thermodynamics characterizes the relation between some physical quantities and the mutual information.
To clarify how the Sagawa-Ueda-Jarzynski equality is modified, we apply our framework to a specific model.
In particular, we deal with a stochastic particle confined in a harmonic potential with a quadratically changing temperature profile.
We emphasize that our generalization could be important when we consider a system that is driven by multiplicative noise and feedback control simultaneously.
Furthermore, we see that our formulation sheds light on conventional stochastic thermodynamics.
That is, by taking the derivative with respect to a temperature profile and the limit of an isothermal system, we can obtain a set of new work relations from the framework. We will exemplify this fact using a specific system.

This paper is organized as follows.
In Sec.~\ref{geo-04-00-01}, we review the Langevin equation of interest and the Sagawa-Ueda-Jarzynski equality.
We then establish an equality for nonuniform temperature systems and discuss its applications in Sec.~\ref{geo-06-01}.
Furthermore, we provide numerical simulation to verify our statement.
In Secs.~\ref{second-law-01} and \ref{geo-06-04}, we investigate applications of the equality.
Finally, Sec.~\ref{conc-08-01} concludes this paper.

%

\section{Sagawa-Ueda-Jarzynski relation} \label{geo-04-00-01}

We here focus on the review of Ref.~\cite{Sagawa03}, in which Sagawa and Ueda generalized the Jarzynski equality by introducing measurement and feedback control.
As explained in Ref.~\cite{Seifert01}, there are mainly three different descriptions for stochastic systems: the Langevin equation, the Fokker-Planck equation, and the Feynman path integral formulation.
Throughout this paper, we mainly use the Langevin equations to specify systems of interest for later convenience.

Although it is slightly different from the original problem setting in Ref.~\cite{Sagawa03}, we consider the following Langevin equation~\cite{Gernert02, Gernert01}:
\begin{align}
\dot{x} (t) &= \frac{1}{m \gamma} f (x (t); \lambda_\mathrm{fb} (t; x_\mathrm{m} (t))) + \bigg( \frac{2 k_\mathrm{B} T_\mathrm{iso}}{m \gamma} \bigg)^{1/2} \xi (t), \label{model-Langevin-SUJ-01}
\end{align}
where
\begin{align}
f (x (t); \lambda_\mathrm{fb} (t; x_\mathrm{m} (t))) &\coloneqq - \frac{\partial}{\partial x} \phi (x (t), \lambda_\mathrm{fb} (t; x_\mathrm{m} (t))),
\end{align}
$\phi (x (t), \lambda_\mathrm{fb} (t; x_\mathrm{m} (t)))$ is the potential that depends on the protocol $\lambda_\mathrm{fb} (t; x_\mathrm{m} (t))$ in which a particle moves, and $\xi (t)$ satisfies $\Braket{\xi (t)} = 0$ and $\Braket{\xi (t) \xi (t')} = \delta (t - t')$.
Here, $m$, $\gamma$, $k_\mathrm{B}$, and $T_\mathrm{iso}$ are the mass of a particle, its friction, the Boltzmann constant, and the temperature of a system.
We also note that $\lambda_\mathrm{fb} (t; x_\mathrm{m} (t)))$ is the protocol depending on $x_\mathrm{m} (t)$, which is a memory based on measurement.
Mathematically, the memory can be rewritten as
\begin{align}
x_\mathrm{m} (t) &=
\begin{cases}
x (t_\mathrm{m}) + B_\mathrm{m} \xi_\mathrm{m} (t_\mathrm{m}) & (t \ge t_\mathrm{m}), \\
x^0_\mathrm{m} & (t < t_\mathrm{m}),
\end{cases} \label{measurement-01}
\end{align}
where $x^0_\mathrm{m}$ is the initial value, and $t_\mathrm{m}$ is the time of a measurement.
Though there is no limitation on the distribution of $\xi_\mathrm{m} (t_\mathrm{m})$, we often assume that $\xi_\mathrm{m} (t_\mathrm{m})$ is drawn from the Gaussian distribution.
For example, when $\xi_\mathrm{m} (t_\mathrm{m})$ is drawn from the Gaussian distribution whose mean is zero and variance is unity, that is, $\xi_\mathrm{m} \sim \mathcal{N} (\cdot; 0, 1)$, then we have
\begin{align}
p (x_\mathrm{m} (t_\mathrm{m}) | x (t_\mathrm{m})) &= \mathcal{N} (x_\mathrm{m} (t_\mathrm{m}); x (t_\mathrm{m}), B_\mathrm{m}^2),
\end{align}
where $\mathcal{N} (x; \mu, \sigma^2)$ is the Gaussian distribution whose mean and variance are, respectively, $\mu$ and $\sigma^2$ on $x$.
Throughout this paper, we ignore the dynamical effect of the memory and consider only a single measurement for simplicity.
The generalizations on the dynamical effect of the memory and multiple measurements are almost straightforward~\cite{Ito01}.

By considering the above system, Sagawa and Ueda showed the following equality in Ref.~\cite{Sagawa03}:
\begin{align}
\Braket{e^{ - \sigma - I^\mathrm{pmi}}} &= 1, \label{equality-Sagawa-21}
\end{align}
where $\sigma \coloneqq \beta (W - \Delta F)$, $W$ is the work performed to the system, $\Delta F$ is the change of the free energy during the given process, $\beta \coloneqq (k_\mathrm{B} T_\mathrm{iso})^{-1}$, and $I^\mathrm{pmi}$ is the pairwise mutual information given by
\begin{align}
I^\mathrm{pmi} &= \ln \frac{p (x (t_\mathrm{m}), x_\mathrm{m} (t_\mathrm{m}))}{p (x (t_\mathrm{m})) p (x_\mathrm{m} (t_\mathrm{m}))}. \label{mutual-info-01}
\end{align}
Note that the conventional mutual information is given by $I^\mathrm{mi} \coloneqq \Braket{I^\mathrm{pmi}}$.
Throughout this paper, the bracket $\Braket{\cdot}$ represents a corresponding expectation value; in the case of Eq.~\eqref{equality-Sagawa-21}, it is given by $\Braket{\cdot} \coloneqq \int \mathcal{D} \Gamma_\mathrm{F} p (\Gamma_\mathrm{F}) [\cdot]$ where $\Gamma_\mathrm{F} \coloneqq \{ x (t) | t: t_\mathrm{ini} \to t_\mathrm{fin} \}$ denotes the trajectory of a forward process.

We note that Eq.~\eqref{equality-Sagawa-21} is the generalization of the Jarzynski equality~\cite{Jarzynski01} under measurement and feedback control and often called the Sagawa-Ueda-Jarzynski relation.
Invoking the Jansen inequality, Sagawa and Ueda obtained that the second law of information thermodynamics~\cite{Sagawa01, Sagawa03, Sagawa05}; furthermore, its validity was confirmed experimentally~\cite{Toyabe01}.

%

\section{Work relations for Tsallis statistical mechanics with measurement and feedback control} \label{geo-06-01}

We derive the Sagawa-Ueda-Jarzynski equality for a nonisothermal system by following Refs.~\cite{Ford01}.
We first construct a general framework.
Then, we apply our framework to a specific model which has a quadratic potential and a quadratic temperature profile discussed in Ref.~\cite{Ford01}.


\subsection{General case} \label{general-case-01}

We consider a feedback control system under a spatially varying temperature profile $T_\mathrm{ni} (x)$.
That is, the system is described by
\begin{align}
\dot{x} (t) &= \frac{1}{m \gamma} f (x (t); \lambda_\mathrm{fb} (t; x_\mathrm{m} (t))) \nonumber \\
& \quad + \bigg( \frac{2 k_\mathrm{B} T_\mathrm{ni} (x (t))}{m \gamma} \bigg)^{1/2} \xi (t). \label{model-g-SUJ-01}
\end{align}
The key point of Eq.~\eqref{model-g-SUJ-01} is that it has the mechanism of feedback control in Eq.~\eqref{model-Langevin-SUJ-01} and $T_\mathrm{ni} (x)$, which is noise that depends on $x$.
The corresponding stochastic equation of Eq.~\eqref{model-g-SUJ-01} is then given by
\begin{align}
dx (t) &= \frac{1}{m \gamma} f (x (t); \lambda_\mathrm{fb} (t; x_\mathrm{m} (t))) dt \nonumber \\
& \quad + \bigg( \frac{2 k_\mathrm{B} T_\mathrm{ni} (x (t))}{m \gamma} \bigg)^{1/2} d \xi (t). \label{model-g-SUJ-02}
\end{align}

By considering the homeomorphic function $f_y (x): x (t) \to y (t)$, we derive an equivalent isothermal system to Eqs.~\eqref{model-g-SUJ-01} and \eqref{model-g-SUJ-02} as follows:
\begin{align}
\dot{y} (t) &= \frac{1}{m \gamma} f_\mathrm{eff} (y (t); \lambda_\mathrm{fb} (t; x_\mathrm{m} (t))) + \bigg( \frac{2 k_\mathrm{B} T_\mathrm{eff}}{m \gamma} \bigg)^{1/2} \xi (t), \label{ito-transform-01}
\end{align}
where
\begin{align}
f_\mathrm{eff} (y (t); \lambda_\mathrm{fb} (t; x_\mathrm{m} (t))) &\coloneqq - \frac{\partial}{\partial y} \phi_\mathrm{eff} (y (t); \lambda_\mathrm{fb} (t; x_\mathrm{m} (t))).
\end{align}
Here, we note that the force and potential in Eq.~\eqref{ito-transform-01} are replaced by the effective ones that will be given later.
Then the equivalent stochastic equation with Eq.~\eqref{ito-transform-01} is written as
\begin{align}
dy (t) &= \frac{1}{m \gamma} f_\mathrm{eff} (y (t); \lambda_\mathrm{fb} (t; x_\mathrm{m} (t))) dt + \bigg( \frac{2 k_\mathrm{B} T_\mathrm{eff}}{m \gamma} \bigg)^{1/2} d \xi (t). \label{system-target-01}
\end{align}
By applying the It\={o} formula~\cite{Oksendal01} to Eq.~\eqref{model-g-SUJ-02}, we obtain
\begin{align}
dy (t) &= \bigg[ \frac{\partial y}{\partial x} \bigg( \frac{1}{m \gamma} f (x (t); \lambda_\mathrm{fb} (t; x_\mathrm{m} (t))) \bigg) \nonumber \\
& \qquad + \frac{1}{2} \frac{\partial^2 y}{\partial x^2} \bigg( \frac{2 k_\mathrm{B} T_\mathrm{ni} (x (t))}{m \gamma} \bigg) \bigg] dt \nonumber \\
& \quad + \frac{\partial y}{\partial x} \bigg( \frac{2 k_\mathrm{B} T_\mathrm{ni} (x (t))}{m \gamma} \bigg)^{1/2} d \xi (t). \label{Ito-rule-04}
\end{align}
Note that we have used $(dt)^2 = 0$ and $(d \xi (t))^2 = dt$ where $d \xi (t)$ represents the white noise whose covariance is unity.
To obtain an isothermal system that has the same form with Eq.~\eqref{system-target-01}, we consider the conditions given by
\begin{align}
\frac{\partial y}{\partial x} &= \bigg( \frac{T_\mathrm{eff}}{T_\mathrm{ni} (x)} \bigg)^{1/2}, \\
\frac{\partial^2 y}{\partial x^2} &= - \frac{T_\mathrm{eff}^{1/2}}{2 (T_\mathrm{ni} (x))^{3/2}} \frac{\partial T_\mathrm{ni} (x)}{\partial x};
\end{align}
thus, we have
\begin{align}
dy (t) &= \bigg( \frac{1}{m \gamma} f (x (t); \lambda_\mathrm{fb} (t; x_\mathrm{m} (t))) \nonumber \\
& \quad + \frac{k_\mathrm{B}}{2 m \gamma} \frac{\partial T_\mathrm{ni} (x (t))}{\partial x} \bigg) \bigg( \frac{T_\mathrm{eff}}{T_\mathrm{ni} (x (t))} \bigg)^{1/2} dt \nonumber \\
& \quad + \bigg( \frac{2 k_\mathrm{B} T_\mathrm{eff}}{m \gamma} \bigg)^{1/2} d \xi (t). \label{equiv-01}
\end{align}
The above expression, Eq.~\eqref{equiv-01}, is an isothermal system equivalent to Eq.~\eqref{model-g-SUJ-02}.

Next, we define the effective work and free energy difference that are consistent with the above transformation.
Then the effective potential for this system is
\begin{align}
& \phi_\mathrm{eff} (y (t); \lambda_\mathrm{fb} (t; x_\mathrm{m} (t))) \nonumber \\
& \quad = \int_0^{y (t)} dy' \, \bigg( \frac{\partial}{\partial x'} \phi (x'; \lambda_\mathrm{fb} (t; x_\mathrm{m} (t))) \nonumber \\
& \qquad + \frac{k_\mathrm{B}}{2 m \gamma} \frac{\partial T_\mathrm{ni} (x')}{\partial x'} \bigg) \bigg( \frac{T_\mathrm{eff}}{T_\mathrm{ni} (x')} \bigg)^{1/2} \nonumber \\
& \quad = \int_0^{x (t) = f_y^{-1} (y(t))} dx' \, \bigg( \frac{\partial}{\partial x'} \phi (x'; \lambda_\mathrm{fb} (t; x_\mathrm{m} (t))) \nonumber \\
& \qquad + \frac{k_\mathrm{B}}{2 m \gamma} \frac{\partial T_\mathrm{ni} (x')}{\partial x'} \bigg) \bigg( \frac{T_\mathrm{eff}}{T_\mathrm{ni} (x')} \bigg). \label{effective-potential-12}
\end{align}
Note that $x'$ and $y'$ are connected via $y' = f_y (x')$.
The effective work for this system is given by
\begin{align}
& W_\mathrm{eff} (y (\cdot), x_\mathrm{m} (\cdot)) \nonumber \\
& \quad = \int_{t_\mathrm{ini}}^{t_\mathrm{fin}} dt \, \frac{\partial}{\partial t'} \phi_\mathrm{eff} (y (t); \lambda_\mathrm{fb} (t'; x_\mathrm{m} (t'))) \bigg|_{t' = t} \label{effect-work-11} \\
& \quad = \int_{t_\mathrm{ini}}^{t_\mathrm{fin}} dt \, \frac{\partial}{\partial t'} \int_0^{x (t) = f_y^{-1} (y(t))} dx' \, \nonumber \\
& \qquad \times \bigg( \frac{\partial}{\partial x'} \phi (x'; \lambda_\mathrm{fb} (t'; x_\mathrm{m} (t'))) \bigg) \bigg( \frac{T_\mathrm{eff}}{T_\mathrm{ni} (x')} \bigg) \bigg|_{t' = t}.
\end{align}
Here, we described $W_\mathrm{eff}$ by using $y (\cdot)$.
In the original coordinate $x (\cdot)$, $W_\mathrm{eff}$ is simply written as
\begin{align}
& W_\mathrm{eff} (x (\cdot), x_\mathrm{m} (\cdot)) \nonumber \\
& \quad = \int_{t_\mathrm{ini}}^{t_\mathrm{fin}} dt \, \frac{\partial}{\partial t'} \int_0^{x (t)} dx' \, \nonumber \\
& \qquad \times \bigg( \frac{\partial}{\partial x'} \phi (x'; \lambda_\mathrm{fb} (t'; x_\mathrm{m} (t'))) \bigg) \bigg( \frac{T_\mathrm{eff}}{T_\mathrm{ni} (x')} \bigg) \bigg|_{t' = t}. \label{effect-work-21}
\end{align}
Note that the mechanical work is given by
\begin{align}
& W (x (\cdot), x_\mathrm{m} (\cdot)) \nonumber \\
& \quad = \int_{t_\mathrm{ini}}^{t_\mathrm{fin}} dt \, \frac{\partial}{\partial t'} \int_0^{x (t)} dx' \, \nonumber \\
& \qquad \times \bigg( \frac{\partial}{\partial x'} \phi (x'; \lambda_\mathrm{fb} (t'; x_\mathrm{m} (t'))) \bigg) \bigg|_{t' = t} \\
& \quad = \int_{t_\mathrm{ini}}^{t_\mathrm{fin}} dt \, \frac{\partial}{\partial t'} \phi (x (t); \lambda_\mathrm{fb} (t'; x_\mathrm{m} (t'))) \bigg|_{t' = t}.
\end{align}
Then, the difference is the modification associated with the nonuniformity of temperature.

The difference of the free energy is give by
\begin{align}
\Delta F_\mathrm{eff} &\coloneqq F_\mathrm{eff} (\phi_\mathrm{eff} (y (t_\mathrm{fin}); \lambda_\mathrm{fb} (t_\mathrm{fin}; x_\mathrm{m} (t_\mathrm{fin})))) \nonumber \\
& \quad - F_\mathrm{eff} (\phi_\mathrm{eff} (y (t_\mathrm{ini}); \lambda_\mathrm{fb} (t_\mathrm{ini}; x_\mathrm{m} (t_\mathrm{ini})))),
\end{align}
where, $F_\mathrm{eff}$ is given by
\begin{align}
F_\mathrm{eff} (\phi_\mathrm{eff} (y; \lambda)) &\coloneqq - \frac{1}{\beta_\mathrm{eff}} \ln \int dy \, \exp (- \beta_\mathrm{eff} \phi_\mathrm{eff} (y; \lambda)). \label{def-eff-free-energy-01}
\end{align}

Therefore, the Sagawa-Ueda-Jarzynski equality for this system is given by
\begin{align}
\Braket{ e^{- \sigma_\mathrm{eff} - I_\mathrm{eff}^\mathrm{pmi}}} &= 1, \label{equality-gene-11}
\end{align}
where
\begin{align}
\sigma_\mathrm{eff} &\coloneqq \beta_\mathrm{eff} (W_\mathrm{eff} - \Delta F_\mathrm{eff}), \label{def-sigma-01} \\
I_\mathrm{eff}^\mathrm{pmi} &\coloneqq \ln \frac{p(y_\mathrm{m} (t_\mathrm{m}), y (t_\mathrm{m}))}{p(y_\mathrm{m} (t_\mathrm{m})) p (y (t_\mathrm{m}))}, \label{def-eff-mutual-01}
\end{align}
with $y_\mathrm{m} \coloneqq f_y (x_\mathrm{m})$.
We have thus shown the general framework of the Sagawa-Ueda-Jarzynski equality for a nonisothermal system.
We note that $\Delta F_\mathrm{eff}$ depends on $x_\mathrm{m} (t)$, and then its expectation value must be taken in Eq.~\eqref{equality-gene-11}.
This is one of the most important properties when we consider measurement explicitly.

%

\subsection{Example I} \label{example-01}

Here, we provide a simple example to illustrate how the general formula, Eq.~\eqref{equality-gene-11}, is applied.
Let us consider a nonisothermal system whose potential and temperature profile are, respectively, given by
\begin{align}
\phi (x (t); \lambda_\mathrm{fb} (t; x_\mathrm{m} (t))) &= \frac{\lambda_\mathrm{fb} (t; x_\mathrm{m} (t))}{2} x^2 (t), \\
T_\mathrm{ni} (x (t)) &= T_\mathrm{eff} \bigg( 1 + \frac{\kappa_T x^2 (t)}{2 k_\mathrm{B} T_\mathrm{eff}} \bigg), \label{temperature-nonisothermal-01}
\end{align}
where $\kappa_T$ is the parameter that specifies the form of the temperature profile.
We note that this system is the same as the model studied in Ref.~\cite{Ford01}.
But the key point is that, in our study, the protocol $\lambda_\mathrm{fb} (t; x_\mathrm{m} (t))$ depends on the measurement $x_\mathrm{m} (t)$ given by Eq.~\eqref{measurement-01}, and thus as we see later some equations are modified by using the measurement and mutual information.

First, the effective potential, Eq.~\eqref{effective-potential-12}, for this system is computed as
\begin{align}
& \phi_\mathrm{eff} (y (t); \lambda_\mathrm{fb} (t; x_\mathrm{m} (t))) \nonumber \\
& \quad = \int_0^{x (t) = f_y^{-1} (y (t))} dx' \, \bigg( \lambda_\mathrm{fb} (t; x_\mathrm{m} (t)) + \frac{1}{2} \kappa_T \bigg) x' \nonumber \\
& \qquad \times \bigg( 1 + \frac{\kappa_T x'^2}{2 k_\mathrm{B} T_\mathrm{eff}} \bigg)^{-1} \nonumber \\
& \quad = \bigg( \lambda_\mathrm{fb} (t; x_\mathrm{m} (t)) + \frac{1}{2} \kappa_T \bigg) \frac{k_\mathrm{B} T_\mathrm{eff}}{\kappa_T} \ln \bigg( 1 + \frac{\kappa_T x^2 (t)}{2 k_\mathrm{B} T_\mathrm{eff}} \bigg).
\end{align}
Then the effective work, Eq.~\eqref{effect-work-11}, for this system is written as
\begin{align}
W_\mathrm{eff} &= \int dt \, \bigg( \frac{\partial}{\partial t'} \lambda_\mathrm{fb} (t'; x_\mathrm{m} (t')) \bigg|_{t' = t} \bigg) \nonumber \\
& \quad \times \frac{k_\mathrm{B} T_\mathrm{eff}}{\kappa_T} \ln \bigg( 1 + \frac{\kappa_T x^2 (t)}{2 k_\mathrm{B} T_\mathrm{eff}} \bigg);
\end{align}
as a result, we have
\begin{align}
\beta_\mathrm{eff} W_\mathrm{eff} &= \int dt \, \dot{W}_\mathrm{eff} \frac{2}{\kappa_T x^2 (t)} \ln \bigg( 1 + \frac{\kappa_T x^2 (t)}{2 k_\mathrm{B} T_\mathrm{eff}} \bigg) \nonumber \\
&= \int dt \, \frac{\dot{W}_\mathrm{eff}}{k_\mathrm{B} T_\mathrm{eff}} \frac{T_\mathrm{eff}}{T_\mathrm{ni} (x (t)) - T_\mathrm{eff}} \ln \frac{T_\mathrm{ni} (x (t))}{T_\mathrm{eff}},
\end{align}
where $\dot{W}_\mathrm{eff} \coloneqq \frac{1}{2} \big( \frac{\partial}{\partial t'} \lambda_\mathrm{fb} (t'; x_\mathrm{m} (t')) \big|_{t' = t} \big) x^2 (t)$.

Next, we consider the free energy difference between $t_\mathrm{ini}$ and $t_\mathrm{fin}$.
To this end, we first compute the stationary distribution at $t_\mathrm{ini}$ and $t_\mathrm{fin}$.
The Fokker-Planck equation equivalent to Eq.~\eqref{model-g-SUJ-01} is given by
\begin{align}
\frac{\partial}{\partial t} p (x, t) &= \frac{\partial}{\partial x} \bigg( \frac{\lambda_\mathrm{fb} (t; x_\mathrm{m} (t)) x}{m \gamma} p (x, t) \nonumber \\
& \quad + \frac{k_\mathrm{B}}{m \gamma} \frac{\partial}{\partial x} \Big[ T_\mathrm{ni} (x) p (x, t) \Big] \bigg).
\end{align}
By adopting the Ito integral for the Langevin equation~\cite{Matsuo01}, we can derive its stationary distribution as follows:
\begin{align}
p_\mathrm{st} (x) &= \bigg( \frac{\kappa_T}{2 \pi k_\mathrm{B} T_\mathrm{eff}} \bigg)^{1/2} \frac{\Gamma (1 + \lambda_\mathrm{fb} (t; x_\mathrm{m} (t)) / \kappa_T)}{\Gamma (1/2 + \lambda_\mathrm{fb} (t; x_\mathrm{m} (t)) / \kappa_T)} \nonumber \\
& \quad \times \bigg( 1 + \frac{\kappa_T x^2}{2 k_\mathrm{B} T_\mathrm{eff}} \bigg)^{- \lambda_\mathrm{fb} (t; x_\mathrm{m} (t)) / \kappa_T - 1}. \label{stationary-dist-11}
\end{align}
We thus can compute the free energy using the stationary distribution~\eqref{stationary-dist-11} as follows:
\begin{align}
& e^{- \beta_\mathrm{eff} F_\mathrm{eff} (\phi_\mathrm{eff} (y (t); \lambda_\mathrm{fb} (t; x_\mathrm{m} (t))))} \nonumber \\
& \quad \propto \int dy \, e^{- \beta_\mathrm{eff} \phi_\mathrm{eff} (y (t); \lambda_\mathrm{fb} (t; x_\mathrm{m} (t)))} \nonumber \\
& \quad = \int dy \, \bigg[ \cosh \bigg( \bigg[ \frac{\kappa_T}{2 k_\mathrm{B} T_\mathrm{eff}} \bigg]^{1/2} y \bigg) \bigg]^{- 2 \frac{\lambda_\mathrm{fb} (t; x_\mathrm{m} (t))}{\kappa_T} -1} \nonumber \\
& \quad = \pi^{1/2} \frac{\Gamma (1/2 + \lambda_\mathrm{fb} (t; x_\mathrm{m} (t)) / \kappa_T)}{\Gamma (1 + \lambda_\mathrm{fb} (t; x_\mathrm{m} (t)) / \kappa_T)}.
\end{align}
Note that $x (t) = \big( \frac{2 k_\mathrm{B} T_\mathrm{eff}}{\kappa_T} \big)^{1/2} \sinh \Big( \big(\frac{\kappa_T}{2 k_\mathrm{B} T_\mathrm{eff}} \big)^{1/2} y (t) \Big)$.
Thus, the free energy difference is computed as
\begin{align}
\Braket{e^{- \beta_\mathrm{eff} \Delta F_\mathrm{eff}}} & = \Braket{\frac{\Gamma (1/2 + \lambda_\mathrm{fb} (t_\mathrm{fin}; x_\mathrm{m} (t_\mathrm{fin})) / \kappa_T)}{\Gamma (1 + \lambda_\mathrm{fb} (t_\mathrm{fin}; x_\mathrm{m} (t_\mathrm{fin})) / \kappa_T)}} \nonumber \\
& \quad \times \frac{\Gamma (1 + \lambda_\mathrm{fb} (t_\mathrm{ini}; x_\mathrm{m} (t_\mathrm{ini})) / \kappa_T)}{\Gamma (1/2 + \lambda_\mathrm{fb} (t_\mathrm{ini}; x_\mathrm{m} (t_\mathrm{ini})) / \kappa_T)}.
\end{align}
Therefore, the Sagawa-Ueda-Jarzynski equality~\eqref{equality-gene-11} for this system is written as
\begin{align}
&\Braket{e^{- \int_{t_\mathrm{ini}}^{t_\mathrm{fin}} dt \, \frac{\dot{W}_\mathrm{eff}}{k_\mathrm{B} T_\mathrm{eff}} \frac{T_\mathrm{eff}}{T_\mathrm{ni} (x) - T_\mathrm{eff}} \ln \frac{T_\mathrm{ni} (x)}{T_\mathrm{eff}} } e^{- I_\mathrm{eff}^\mathrm{pmi}}} \nonumber \\
& \quad = \Braket{\frac{\Gamma (1/2 + \lambda_\mathrm{fb} (t_\mathrm{fin}; x_\mathrm{m} (t_\mathrm{fin})) / \kappa_T)}{\Gamma (1 + \lambda_\mathrm{fb} (t_\mathrm{fin}; x_\mathrm{m} (t_\mathrm{fin})) / \kappa_T)}} \nonumber \\
& \qquad \times \frac{\Gamma (1 + \lambda_\mathrm{fb} (t_\mathrm{ini}; x_\mathrm{m} (t_\mathrm{ini})) / \kappa_T)}{\Gamma (1/2 + \lambda_\mathrm{fb} (t_\mathrm{ini}; x_\mathrm{m} (t_\mathrm{ini})) / \kappa_T)}. \label{generalized-relaiton-example-01}
\end{align}

Finally, let us remark on the system that we have dealt with.
The stationary distribution, Eq.~\eqref{stationary-dist-11}, is also written as the extended exponential defined by Tsallis~\cite{Tsallis03, Tsallis01}; that is,
\begin{align}
p_\mathrm{st} (x) &\propto e_q (z),
\end{align}
where
\begin{align}
e_q (z) &= [1 + (1 - q) z]^{1 / (1 - q)}, \\
q &= \frac{\lambda_\mathrm{fb} (t; x_\mathrm{m} (t)) + 2 \kappa_T}{\lambda_\mathrm{fb} (t; x_\mathrm{m} (t)) + \kappa_T}, \\
z &= - \frac{(\lambda_\mathrm{fb} (t; x_\mathrm{m} (t)) + \kappa_T) x^2}{2 k_\mathrm{B} T_\mathrm{eff}}.
\end{align}
Thus, our extension can be interpreted as an extension of the Sagawa-Ueda-Jarzynski equality based on Tsallis statistical mechanics.

\subsection{Example II} \label{exmaple-03-01}

To perform a numerical simulation, we consider a simple process as follows.
First, we consider the potential given
\begin{align}
\phi (x) &= \frac{1}{2} \lambda (t) x^2,
\end{align}
where
\begin{align}
\lambda (t) &= \lambda_i & (t_i \le t \le t_{i+1}).
\end{align}
This process is sometimes called the cycle process because the forms of the potentials at the initial and final states are the same; as a result the change of the free energy $\Delta F_\mathrm{eff}$ is zero.
In this case, we have
\begin{align}
\beta_\mathrm{eff} W_\mathrm{eff} &= \sum_{i=0}^2 \frac{\lambda_{i+1} - \lambda_i}{\kappa_T} \ln \bigg(1 + \frac{\kappa_T x(t_i)^2}{2 k_\mathrm{B} T_\mathrm{eff}} \bigg). \label{numerical-01}
\end{align}

In the numerical simulation, we set $k_\mathrm{B} = 1$, $T_\mathrm{eff} = 1$, $\lambda_0 = \lambda_2 = 1$, $\lambda_1 = 2 (1 + x_\mathrm{m}(t_\mathrm{m})^2)$, $t_0 = 0$, $t_1 =1$, $t_2 = 2$, and $t_3 = 3$.
We measure the state $x (t_\mathrm{m})$ at $t_\mathrm{m} = t_1$ and the measurement error is assumed to be standard white noise; that is, $x_\mathrm{m} (t_\mathrm{m}) \sim \mathcal{N} (x_\mathrm{m} (t_\mathrm{m}); x (t_\mathrm{m}), 1.0)$.

\begin{figure}[t]
\centering
\includegraphics[scale=0.45]{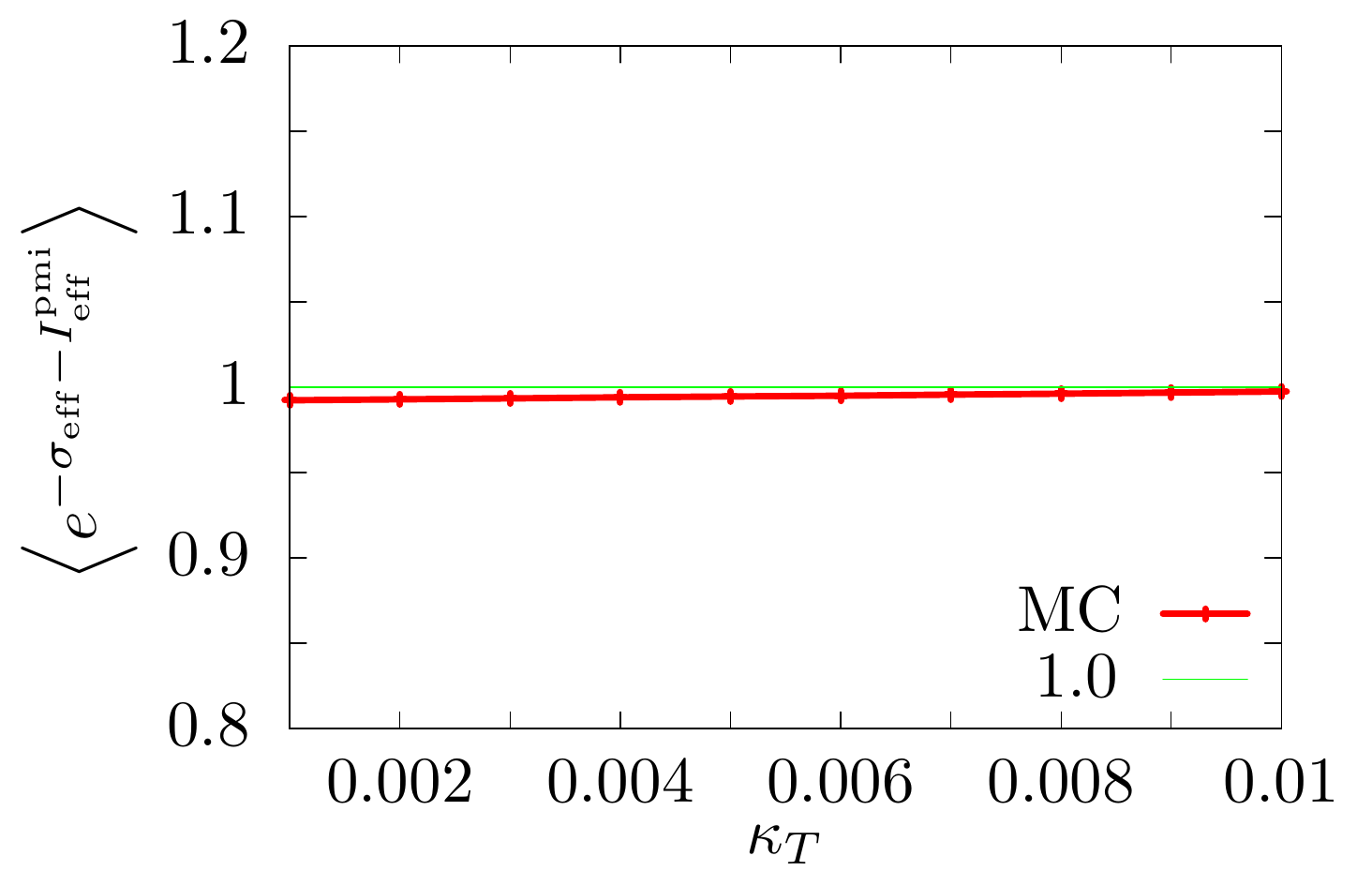}
\caption{Numerical results of Eq.~\eqref{equality-gene-11} for $\kappa_T = 0.001, 0.002, \dots, 0.01$. The vertical axis is dimensionless. We set $k_\mathrm{B} = 1$, $T_\mathrm{eff} = 1$, $\lambda_0 = \lambda_2 = 1$, $\lambda_1 = 2 (1 + x_\mathrm{m}(t_\mathrm{m})^2)$, $t_0 = 0$, $t_1 =1$, $t_2 = 2$, and $t_3 = 3$.}
\label{numerical-01-01}
\end{figure}

We performed the process $10^7$ times to compute the ensemble average $\Braket{\cdot}$ and obtained $0.997 \pm 0.001$ for the left hand side of Eq.~\eqref{numerical-01}.
In addition, we varied $\kappa_T = 0.001, 0.002, \dots, 0.01$ and summarized the results in Fig.~\ref{numerical-01-01}.
These numerical simulations support the validity of one of our main claims, Eq.~\eqref{equality-gene-11}.

%


\section{Generalized second law of information thermodynamics} \label{second-law-01}

We consider generalizing the second law of information thermodynamics~\cite{Sagawa01, Sagawa03, Sagawa05}.
By employing the concavity of the exponential function, Eq.~\eqref{equality-gene-11} yields
\begin{align}
\beta_\mathrm{eff} \Braket{W_\mathrm{eff}} - \beta_\mathrm{eff} \Braket{\Delta F_\mathrm{eff}} + \Braket{I_\mathrm{eff}^\mathrm{pmi}} &\ge 0. \label{gene-second-01}
\end{align}
This is the second law of information thermodynamics for a nonisothermal system.
We mention a property concerning the mutual information.
When the change of variables $y = f_y (x)$ is homeomorphic, that is, continuous and uniquely invertible, mutual information is invariant under the transformation; that is, $\Braket{I_\mathrm{eff}^\mathrm{pmi}} = \Braket{I^\mathrm{pmi}}$ holds~\cite{Kraskov01} (see Sec.~\ref{app-11-01}).

We have derived the inequality, Eq.~\eqref{gene-second-01}; then, it is natural to consider the condition for the equality.
The inequality derived from Eq.~\eqref{equality-Sagawa-21} provides an equality in the quasi-static process.
Thus, if we can derive an optimal protocol, Eq.~\eqref{gene-second-01} is expected to be an equality in the case of the quasi-static process.
But in general it is difficult to derive an optimal protocol because as demonstrated in Ref.~\cite{Abreu01} we have to solve a nonlinear equation that is defined via a stationary distribution.
Furthermore, it is also difficult to show that such an optimal protocol satisfies the equality.
Note that, by ignoring the third term of the left-hand side, Eq.~\eqref{gene-second-01} reduces to the generalized second law of thermodynamics concerning a nonisothermal system.

%

\section{New relations for an isothermal system} \label{geo-06-04}

We mainly consider a nonisothermal system in this paper; however, our formulation provides a set of new work relations on an isothermal system at $T_\mathrm{eff}$.
For simplicity, we consider that the nonuniformity of $T_\mathrm{ni} (x)$ is characterized by a single parameter $\kappa_T$; that is, when $\kappa_T = 0$, then $T_\mathrm{ni} (x)$ does not depend on $x$.
A typical example is Eq.~\eqref{temperature-nonisothermal-01}.

We define $a (\kappa_T)$ and $b (\kappa_T)$ by
\begin{align}
  a (\kappa_T) &= \Braket{e^{- \beta_\mathrm{eff} W_\mathrm{eff} - I_\mathrm{eff}^\mathrm{pmi}}}, \\
  b (\kappa_T) &= \Braket{e^{- \beta_\mathrm{eff} \Delta F}},
\end{align}
respectively.
Then Eq.~\eqref{equality-gene-11} is rewritten as $a (\kappa_T) = b (\kappa_T)$.
Computing the derivatives with respect to $\kappa_T$, we have
\begin{align}
  \frac{\partial^n}{\partial \kappa_T^n} a (\kappa_T) &= \frac{\partial^n}{\partial \kappa_T^n} b (\kappa_T),
\end{align}
for each integer $n$.
Furthermore, taking the limit $\kappa_T \to 0$, we obtain
\begin{align}
  \lim_{\kappa_T \to 0} \frac{\partial^n}{\partial \kappa_T^n} a (\kappa_T) &= \lim_{\kappa_T \to 0} \frac{\partial^n}{\partial \kappa_T^n} b (\kappa_T), \label{new-equality-boltzmann-01}
\end{align}
We note that Eq.~\eqref{new-equality-boltzmann-01} provides a set of new work relations on an isothermal system at $T_\mathrm{eff}$.

%

\section{Conclusion} \label{conc-08-01}

We have investigated a nonisothermal system with measurement and feedback control, and derived the generalized Sagawa-Ueda-Jarzynski equality.
Moreover, we have applied our framework to a specific model and provided a concrete expression.
Then, we have discussed the generalized second law of information thermodynamics.
We expect that our contributions can be applied to various kinds of systems including mesoscopic systems and biological systems, and the validity can be discussed.
Finally, we have also derived a set of new work relations for isothermal systems.

%

\section*{Acknowledgements} \label{ack-08-01}

The authors thank to Takahiro Sagawa for reading our first manuscript and providing fruitful comments.
This research is supported by JSPS KAKENHI, Grant No. JP18J12175, and JST CREST, Grand No. JPMJCR14D2, Japan.

%


\appendix

\section{Homeomorphic function} \label{app-11-01}

In Ref.~\cite{Kraskov01}, it is addressed that the mutual information is unchanged under a transformation that is homeomorphic (continuos and uniquely invertible maps).

To clarify the statement, let random variables $X$ and $Y$ be drawn from $p (x)$ and $p (y)$, respectively; that is,
\begin{align}
X &\sim p (x), \\
Y &\sim p (y).
\end{align}
Furthermore, we consider two homeomorphic functions $F (\cdot)$ and $G (\cdot)$:
\begin{align}
X' &= F(X), \\
Y' &= G(Y).
\end{align}
Letting $J_F (x')$ and $J_G (y')$ be the Jacobian matirces, the measure function on $x'$ and $y'$ is described as
\begin{align}
\mu_{x', y'} (x', y') &= J_F (x') J_G (y') \mu_{x, y} (x, y).
\end{align}
Similarly, we have
\begin{align}
\mu_{x'} (x') &= J_F (x') \mu_x (x), \\
\mu_{y'} (y') &= J_F (y') \mu_y (y).
\end{align}

The mutual information between $X$ and $Y$ is given by
\begin{align}
I (X; Y) &= \int dx \, dy \, \mu_{x, y} (x, y) \ln \frac{\mu_{x, y} (x, y)}{\mu_x (x) \mu_y (y)}. \label{def-mutual-info}
\end{align}
Then, the mutual information between $X'$ and $Y'$ is computed as
\begin{align}
I (X'; Y') &= \int dx' \, dy' \, \mu_{x', y'} (x', y') \ln \frac{\mu_{x', y'} (x', y')}{\mu_{x'} (x') \mu_{y'} (y')} \\
&= \int dx \, dy \, \mu_{x, y} (x, y) \ln \frac{\mu_{x, y} (x, y)}{\mu_{x} (x) \mu_{y} (y)} \\
&= I (X; Y).
\end{align}
The above discussion is limited to the logarithmic function $\ln (\cdot)$, which appears in the definition of the mutual information~\eqref{def-mutual-info}, but this discussion can be generalized to a general function $f(\cdot)$.
If we define $I_f (X; Y)$ as
\begin{align}
I_f (X; Y) &= \int dx \, dy \, \mu_{x, y} (x, y) f \bigg( \frac{\mu_{x, y} (x, y)}{\mu_x (x) \mu_y (y)} \bigg),
\end{align}
then we have
\begin{align}
I_f (X'; Y') &= \int dx' \, dy' \, \mu_{x', y'} (x', y') f \bigg( \frac{\mu_{x', y'} (x', y')}{\mu_{x'} (x') \mu_{y'} (y')} \bigg) \\
&= \int dx \, dy \, \mu_{x, y} (x, y) f \bigg( \frac{\mu_{x, y} (x, y)}{\mu_{x} (x) \mu_{y} (y)} \bigg) \\
&= I_f (X; Y). \label{gene-ch-mutual-info}
\end{align}
This formula~\eqref{gene-ch-mutual-info} can be applied to Eq.~\eqref{equality-gene-11} to relpace $I_\mathrm{eff}^\mathrm{pmi}$ with $I^\mathrm{pmi}$, Eq.~\eqref{mutual-info-01}.

\bibliographystyle{apsrev4-1}
\bibliography{paper-gene-stat-mech-99-01-bib}

\end{document}